\documentclass[aps,showpacs]{revtex4}

\usepackage{amssymb}

\newcommand*{\bra}[1]{\langle#1|}
\newcommand*{\ket}[1]{|#1\rangle}
   \newcommand*{\bx}{\mathbf{x}}
   
   \newcommand*{\da}{\dagger}
   \newcommand*{\ep}{\epsilon}

   \newcommand*{\ps}{\psi} 
   \newcommand*{\si}{\sigma} 
   \newcommand*{\De}{\Delta}                                          
   \newcommand*{\Eq}[1]{Eq.~(\ref{eq:#1})}
   \newcommand*{\eq}[1]{(\ref{eq:#1})}
   \newcommand*{\bracket}[1]{\langle#1\rangle}
  \newcommand{\bS}{{\bf S}}
  \newcommand{\cH}{{\mathcal H}}
  \newcommand{\bA}{{\bf A}}
  \newcommand{\by}{{\bf y}}
  \newcommand{\R}{{\bf R}}
  \newcommand{\cM}{{\mathcal M}}
 \newcommand{\cK}{{\mathcal K}}
\newcommand{\braket}[1]{\langle#1\rangle}
 
  \newcommand{\rh}{\rho}
  \newcommand{\Tr}{\mbox{\rm Tr}}
  \newcommand{\tA}{\tilde{A}}
  \newcommand{\tB}{\tilde{B}}
 \newcommand{\tX}{\tilde{X}}

\begin{document}

\title{Uncertainty Relations for Joint Measurements
of Noncommuting Observables}\thanks{
The results in this Letter were presented 
 at A Satellite Workshop to EQIS'03:
Non-locality of Quantum Mechanics and Statistical Inference,  
held at Kyoto Sangyo University, Kyoto, Japan, on September 8--9, 2003.
}
\author{Masanao Ozawa}
\email{ozawa@mailaps.org}
\affiliation{Graduate School of Information Sciences,
T\^{o}hoku University, Aoba-ku, Sendai,  980-8579, Japan}

\begin{abstract}
Universally valid uncertainty relations are proven 
in a model independent formulation for inherent and 
unavoidable extra noises in arbitrary joint measurements 
on single systems,
from which Heisenber's original uncertainty relation is 
proven valid for any joint measurements 
with statistically independent noises. 
\end{abstract}
\pacs{03.65.Ta,  03.67.-a}
\maketitle

\section{Introduction}

Heisenberg's uncertainty principle \cite{Hei27,Hei30} represents a
fundamental limit on measurements in quantum mechanics 
as a consequence of noncommutativity of canonical observables. 
However, its interpretation and formalization have been an issue 
of debate for many years. 

Heisenberg  \cite{Hei27} originally asserted the relation (Heisenberg's
relation) that the product of the
imprecisions of measurements of conjugate observables carried out
simultaneously on a system has a lower bound determined 
by Planck's constant, as exemplified by the $\gamma$ ray microscope 
thought experiment. 
Some of previous investigations, on the one hand, partially supported the
validity of Heisenberg's relation
\cite{AK65,YH86,AG88,91QU,Ish91},  and explained the consistency
of Heisenberg's relation with the sensitivity  limit of experimentally available
optical detectors and amplifiers \cite{SH66,Yue82,YH86,AG88}.
On the other hand, there have been criticisms on the validity of 
Heisenberg's relation and its consequences  
\cite{Yue83,MM90,88MS,89RS}.
Moreover, recent investigations  \cite{01CQSR,02KB5E,03UVR} have
revealed that Heisenberg's relation is violated in a rather common class of
measurements.  

We have also different formalizations of the uncertainty principle
introduced by Kennard \cite{Ken27} and Robertson \cite{Rob29}. 
Robertson's relation for a pair of observables $A$ and
$B$ is formalized by the relation
\begin{equation}\label{eq:Robertson}
\De A\De B\ge \frac{1}{2}|\bracket{[A,B]}|,
\end{equation}
where 
$\bracket{\cdots}$ stands for the mean value in the
given state,
$\De A$ and $\De B$ are the standard deviations of $A$ and $B$,
defined by 
$\De X=(\bracket{X^{2}}-\bracket{X}^{2})^{1/2}$ for $X=A,B$,
and the square bracket stands for the commutator, i.e., $[A,B]=AB-BA$.
Equation \eq{Robertson} can be proven by a simple application of
the Schwarz inequality.  
For two conjugate observables $Q$ and $P$, which satisfy 
the canonical commutation relation
\begin{equation}\label{eq:CCR}
[Q,P]=i\hbar,
\end{equation}
we obtain Kennard's relation \cite{Ken27}
\begin{equation}\label{eq:HUR}
\De Q \De P\ge\frac{\hbar}{2}.
\end{equation}

Robertson's and Kennard's relations are naturally interpreted
as the limitation of state preparations or the limitation of the ideal
independent measurements on identically prepared systems 
\cite{Bal70,Per93}.
Moreover, the standard deviation, a notion dependent on the state of the
system but independent of the apparatus,  
cannot be identified with the imprecision of the apparatus such as the
resolution power of the $\gamma$ ray microscope.
Thus, it is still missing to correctly describe the unavoidable imprecisions 
inherent to joint measurements of noncommuting observables. 

The purpose of this Letter is to present a correct, general, and
relevant formalization of the uncertainty principle
for joint measurements in the model independent formulation 
based on the notion of positive operator valued measures, 
now broadly accepted as the most general  description of 
statistics of measurement outcomes 
\cite{Hel76,Dav76,Hol82,84QC,Per93,NC00}, 
and to show that measurements obeying Heisenberg's 
relation are generally characterized by the statistical  
independence of the noise in measurement. 

\section{Violation of Heisenberg's relation}

Before discussing the general formulation, 
we shall see how Heisenberg's relation is commonly violated. 
The well-known device to realize approximate simultaneous 
measurements of conjugate observables
\cite{AK65,Yue82,YH86} is based on the fact that two 
independent pairs of conjugate  observables $(Q, P)$ 
and $(Q',P')$ satisfy the commutation relation 
\begin{equation}
[Q-Q',P+P']=0,
\end{equation}
where the independence ensures the relation $[Q,P']=[Q',P]=0$.
Thus, we have an apparatus to simultaneously measure 
$Q_{-}=Q-Q'$ and $P_{+}=P+P'$ with arbitrary precision.
The {\em precise} simultaneous measurement of $Q_{-}$ and $P_{+}$
can be interpreted as an {\em approximate} simultaneous
measurement of $Q$ and $P$, if we take the outputs from the 
above measurement to be the measured values of $Q$ and $P$. 
How precisely $Q$ has been measured is evaluated by the difference 
between the observable $Q$ to be measured and the observable $Q_{-}$
actually measured.  Thus, the root-mean-square (rms) noise 
$\ep(Q)=\bracket{(Q_{-}-Q)^{2}}^{1/2}$ is a
reasonable measure of imprecision for the $Q$ measurement.
Analogously, so is the rms noise 
$\ep(P)=\bracket{(P_{+}-P)^{2}}^{1/2}$ 
for the $P$ measurement.
By  $Q_{-}-Q=Q'$, $P_{+}-P=P'$, we have
\begin{eqnarray}
\ep(Q)&=&\bracket{Q'{}^{2}}^{1/2}\ge\De Q',
\label{eq:Q-noise}\\
\ep(P)&=&\bracket{P'{}^{2}}^{1/2}\ge\De P',
\end{eqnarray} 
so that the relation \eq{HUR}, applied to the pair $(Q',P')$,
concludes Heisenberg's relation
\begin{equation}\label{eq:Heisenberg}
\ep(Q)\ep(P)\ge\frac{\hbar}{2}.
\end{equation}
Thus, the simultaneous measurement of commuting observables
$Q_{-}$ and $P_{+}$ can be regarded as an approximate 
simultaneous measurement  of conjugate observables $Q$ and
$P$ which satisfies Heisenberg's relation.

Now, we consider other independent pairs $(Q,P_{+})$ 
and $(Q_{-},-P)$ of conjugate observables
produced by the canonical transformation
$Q\to Q$, $P\to P_{+}$, $Q'\to Q_{-}$, and $P'\to -P$.
Then, the precise simultaneous measurement of $Q_{-}$ and $P_{+}$
is also interpreted as an approximate simultaneous measurement of
conjugate observables $Q$ and $P_{+}$. 
In this case, we obviously have $\ep(P_{+})=0$.
Thus, if Heisenberg's relation were to hold, 
we would have $\ep(Q)=\infty$.
However, if $Q'$ would be prepared to be $Q'=0$,  
the precise $Q_{-}$ measurement could be regarded as the
precise $Q$ measurement.
Indeed, if the mean position of $Q'$ is prepared at the origin,
i.e., $\bracket{Q'}=0$, then from \Eq{Q-noise} we have 
$\ep(Q)=\De Q'$, so that we can make $\ep(Q)$ arbitrarily small,
and we conclude
\begin{equation}\label{eq:violation}
\ep(Q)\ep(P_{+})=0.
\end{equation}
Thus, Heisenberg's relation is violated, and more significantly
we have shown that the position $Q$ can be measured with
arbitrary precision without disturbing the total momentum $P+P'$.

Therefore, we have shown that the precise simultaneous measurement
of commuting observables $Q_{-}$ and $P_{+}$ can be regarded,
on the one hand,
as an approximate simultaneous measurement of conjugate
observables $Q$ and $P$ satisfying Heisenberg's relation 
and, on the other hand, as an effectively precise simultaneous 
measurement of conjugate observables $Q$ and $P_{+}$.
In particular, both \Eq{Heisenberg} and \Eq{violation} hold
mathematically for every state 
with $0<\De Q,\De P,\De Q',\De P'<0$.

This contradicts also  
a naive interpretation of the Wigner-Araki-Yanase
theorem stating that only observables commuting with all
the additively conserved quantities can be measured precisely
\cite{Wig52,AY60,Yan61,Wig63,91CP,02CLU}.
Our generalized uncertainty relations will clarify, among others,
what limitation is generally posed in the above situation.

\section{Joint probability operator valued measures}

Let us consider a quantum system $\bS$ 
described by a Hilbert space $\cH$ 
with two observables $A$ and $B$.
We associate any joint measurement 
of observables $A$ and $B$
to an apparatus $\bA(\bx,\by)$ 
with two output variables $\bx$ and $\by$, 
where $\bx$ measures $A$ and $\by$ measures $B$, 
even approximately.
We assume that any joint measurements are carried out
on a single systems to obtain,
simultaneously or successively in time, 
the measured values of $A$ and $B$ 
in a common state just before the measurement.
For simplicity, we assume that $\cH$ is finite dimensional and
that $\bx$ and $\by$ take only finite number of values.
It is now fairly well-known
\cite{Hel76,Dav76,Hol82,84QC,Per93,NC00} that to
such an apparatus
$\bA$ we can associate a family $\{\Pi(x,y)|\ x,y\in\R\}$ of
positive operators on $\cH$ satisfying (i) $0\le \Pi(x,y)\le I$,
(ii) $\sum_{x,y}\Pi(x,y)=I$.
Then, the joint probability of obtaining the outcomes 
$\bx=x$ and $\by=y$ in the state $\ps$ is described by
\begin{equation}
\Pr\{\bx=x,\by=y\|\ps\}=\bracket{\ps|\Pi(x,y)|\ps}.
\end{equation}
We shall call the above family $\{\Pi(x,y)|\ x,y\in\R\}$
the joint probability operator valued measure (POVM) of the
apparatus $\bA(\bx,\by)$.
The joint POVM $\Pi(x,y)$ defines two marginal 
POVMs $\Pi^{A}(x)$ and $\Pi^{B}(y)$ 
by $\Pi^{A}(x)=\sum_{y}\Pi(x,y)$ and 
$\Pi^{B}(y)=\sum_{x}\Pi(x,y)$.
The marginal POVMs $\Pi^{A}(x)$ and $\Pi^{B}(y)$
describe the output probability distributions for the $A$ 
measurement and the $B$ measurement, respectively, i.e.,
$\Pr\{\bx=x\|\ps\}=\bracket{\ps|\Pi^{A}(x)|\ps}$ and 
$\Pr\{\by=y\|\ps\}=\bracket{\ps|\Pi^{B}(y)|\ps}$.

From the Born rule for the probability distributions of observables, 
the joint POVM $\Pi(x,y)$ gives a precise
$A$ measurement if and only if
$ 
\Pi^{A}(x)=E^{A}(x)
$
for all $x\in\R$, where $E^{A}(x)$ is the projection 
on the subspace $\{\ps\in\cH|\ A\ps=x\ps\}$.
Analogously, $\Pi(x,y)$ gives a precise $B$ measurement
if and only if
$ 
\Pi^{B}(y)=E^{B}(y)
$ 
for all $y\in\R$.
It is well-known that $\Pi(x,y)$ gives both a precise $A$ 
measurement and a precise $B$ measurement if and only if
\begin{equation}
\Pi(x,y)=E^{A}(x)E^{B}(y)
\end{equation}
for all $x,y\in\R$  \cite{Dav76}.  
Since $\Pi(x,y)^{\da}=\Pi(x,y)$, we have
$[E^{A}(x),E^{B}(y)]=0$ for all $x,y\in\R$, so that $[A,B]=0$.
Thus, the precise simultaneous measurement of $A$ and $B$
is possible if and only if $A$ and $B$ commute.
However, the above result does not exclude the possibility 
that there is a subspace $\cM$ such that an effectively precise 
simultaneous measurement of $A$ and $B$  are possible 
for any state in $\cM$ on which $[A,B]\not=0$.
For such problem the quantitative investigation
is inevitable.

\section{Measuring processes}

The above statistical description of measurement by the POVM
is known to be consistent with a description
of measuring process \cite{84QC}.
A measuring process for the joint POVM $\Pi(x,y)$ is defined 
to be a 5-tuple $(\cK,\xi,U,M_{1},M_{2})$ consisting of  
a Hilbert space $\cK$, a state vector $\xi$ on $\cK$,
a unitary operator $U$ on $\cH\otimes\cK$, 
and commuting
observables $M_{1}$ and $M_{2}$ on $\cK$ such that
\begin{equation}\label{eq:joint-MP}
\Pi(x,y)=\bracket{\xi|U^{\da}[I\otimes E^{M_{1}}(x)
E^{M_{2}}(y)]U|\xi}
\end{equation}
\sloppy
for any $x,y\in\R$, where $\bracket{\xi|\cdots|\xi}$
is the partial mean on $\cK$, i.e., 
$\braket{\ps|\bracket{\xi|\cdots|\xi}|\ps}
=\braket{\ps\otimes\xi|\cdots|\ps\otimes\xi}$
for all $\ps\in\cH$.
If the joint measurement described by $\Pi(x,y)$ is carried out 
by a measuring process
$(\cK,\si,U,M_{1},M_{2})$, the input state $\rh$ 
is changed to the conditional output state $\rh_{\{\bx=x,\by=y\}}$
\begin{eqnarray}
\rh_{\{\bx=x,\by=y\}
                  }
&=&
\frac
{\Tr_{\cK}\{U(\rh\otimes\ket{\xi}\bra{\xi})U^{\da}[I\otimes
E^{M_{1}}(x) E^{M_{2}}(y)]\}
}
{\Tr\{U(\rh\otimes\ket{\xi}\bra{\xi})U^{\da}[I\otimes
E^{M_{1}}(x) E^{M_{2}}(y)]\}
},\quad
\end{eqnarray} 
provided that the measurement leads to the outcome 
``$\bx=x$ and $\by=y$''.

In Ref.~ \cite{84QC}, it was proven that for any joint POVM $\Pi(x,y)$
there is at least one measuring process $(\cK,\xi,U,M_{1},M_{2})$
for $\Pi(x,y)$, which satisfies \Eq{joint-MP} for any $x,y\in\R$.
By defining $C=U^{\da}(I\otimes M_{1})U$ and 
$D=U^{\da}(I\otimes M_{2})U$ in \Eq{joint-MP}, we conclude that
for any joint POVM $\Pi(x,y)$ there exist
a Hilbert space $\cK$, a state vector $\xi$
in $\cK$, and commuting observables $C,D$ on $\cH\otimes\cK$ 
such that
\begin{equation}\label{eq:ancilla}
\Pi(x,y)=\bracket{\xi|E^{C}(x)E^{D}(y)|\xi}
\end{equation}
for any $x,y\in\R$.
We generally call any quadruple $(\cK,\xi,C,D)$ satisfying \Eq{ancilla}
for any $x,y\in\R$ the ancilla for
the joint POVM $\Pi(x,y)$.

From \Eq{ancilla}, the joint measurement of $A$ and $B$ in state $\ps$
described by $\Pi(x,y)$ is statistically equivalent to the simultaneous
measurement of commuting observables $C$ and $D$ in the state
$\ps\otimes\xi$. Thus, the ancilla $(\cK,\xi,C,D)$ defines the {\em noise} 
$N_{A}$ in the $A$ measurement by the relation
\begin{equation}\label{eq:030819c}
C=A\otimes I +N_{A}.
\end{equation}
Thus, the noise is the difference between the observable to be measured
and the observable actually measured.
Since the ancilla is always prepared in the state $\xi$, the partial mean
of the noise over the ancilla defines the {\em mean noise operator}
$n_{A}$ for the $A$ measurement by
\begin{equation}
n_{A}=\bracket{\xi|N_{A}|\xi}.
\end{equation}
The {\em rms noise} $\ep(A)$ for the $A$ measurement in the
system state
$\ps$ is naturally defined by 
\begin{equation}
\ep(A)
=\bracket{N_{A}^{2}}^{1/2},
\end{equation}
where $\bracket{\cdots}$ stands for the mean values in the
state $\ps\otimes\xi$.
By an easy computations, we have 
\begin{eqnarray}
n_{A}&=&O(\Pi^{A})-A,\label{eq:output-mean}\\
\ep(A)^{2}&=&\bracket{\ps|O^{(2)}(\Pi^{A})-O(\Pi^{A})^{2}+n_{A}^{2}
|\ps},\\
(\De N_{A})^{2}&=&\ep(A)^{2}
-\bracket{\ps|n_{A}|\ps}^{2},
\end{eqnarray}
where $O(\Pi^{A})$ and $O^{(2)}(\Pi^{A})$ are the first and
the second moment operators defined by 
$O(\Pi^{A})=\sum_{x}\,x\,\Pi^{A}(x)$ and
$O^{2}(\Pi^{A})=\sum_{x}\,x^{2}\,\Pi^{A}(x)$,
and $\De N_{A}$ is the standard deviation of the noise $N_{A}$.
We also have the following expressions.
\begin{eqnarray*}
n_{A}&=&\sum_{x}\Pi^{A}(x)(x-A),\\
\ep(A)^{2}&=&\sum_{x}\|\Pi^{A}(x)^{1/2}(x-A)\ps\|^{2}.
\end{eqnarray*}
The above relations show that the mean noise operator,  
the rms noise, and the standard deviation of noise are
intrinsic properties of the POVM independent of particular
description of the ancilla system.

It can be shown that the POVM $\Pi^{A}$ precisely measures 
an observable $A$, i.e,
$\Pi^{A}(x) =E^{A}(x)$ for any $x$ 
if and only if $\ep(A)=0$  for any state $\ps$ of $A$.
The corresponding formulations for the $B$ measurement 
can be given analogously.

\section{Universally valid uncertainty relations}

Under the above formulation for joint measurements and their noises,
the generalized uncertainty relation is obtained as follows (see 
Refs.~\cite{03UVR,03HUR,03URN} for a parallel argument 
for noise-disturbance uncertainty relations).  Since $[C,D]=0$, we obtain
$[\tA+N_{A},\tB+N_{B}]=0$, where $\tA=A\otimes I$ and $\tB=B\otimes I$.
Then, we have \cite{YH86}
\begin{eqnarray}
[N_{A},N_{B}]\!+\![N_{A},\tB]\!+\![{\tA},N_{B}]\!=\!-[A,B]\!\otimes\! I.\quad
\end{eqnarray}
Taking the moduli of means of the both sides
 and applying the triangular inequality, we have
\begin{eqnarray}\label{eq:commutation2}
        |\bracket{[N_{A},           N_{B}]}|
+|\bracket{[{\tA},N_{B}]}|+|\bracket{[N_{A},{\tB}]}|
\ge |\bracket{\ps|[A,B]|\ps}|.
\end{eqnarray}
From Robertson's relation we have
\begin{equation}\label{eq:Robertson-N}
\ep(A)\ep(B)\ge\De N_{A}\De N_{B}\ge\frac{1}{2}|\bracket{[N_{A},N_{B}]}|.
\end{equation}
By the relations
\begin{equation}\label{eq:3617b}
\bracket{N_{A}\tB}=\bracket{\ps|\bracket{\xi|N_{A}|\xi}|B\ps}=
\bracket{\ps|n_{A}B|\ps},
\end{equation}
 we have
\begin{equation}\label{eq:3618a}
\bracket{[N_{A},{\tB}]}=\bracket{\ps|[n_{A},B]|\ps}.
\end{equation}
Similarly, we also have
\begin{equation}\label{eq:3618b}
\bracket{[{\tA},N_{B}]}=\bracket{\ps|[A,n_{B}]|\ps}.
\end{equation}
Therefore, by substituting Eqs.~\eq{Robertson-N}, \eq{3618a}, and
\eq{3618b} in \Eq{commutation2}, we obtain 
\begin{eqnarray}\label{eq:UVUR3}
\De N_{A}\,\De N_{B}
+\frac{1}{2}|\bracket{\ps|[n_{A},B]|\ps}|
+\frac{1}{2}|\bracket{\ps|[A,n_{B}]|\ps}|
&\ge&\frac{1}{2}|\bracket{\ps|[A,B]|\ps}|,\quad
\end{eqnarray}
and the {\em 
universally valid uncertainty relation for joint measurement} 
\begin{eqnarray}\label{eq:UVUR2}
\ep(A)\ep(B)
+\frac{1}{2}|\bracket{\ps|[n_{A},B]|\ps}|
+\frac{1}{2}|\bracket{\ps|[A,n_{B}]|\ps}|
&\ge&\frac{1}{2}|\bracket{\ps|[A,B]|\ps}|.
\end{eqnarray}

In order to obtain the trade-off among the rms noises
$\ep(A)$ and $\ep(B)$, 
and the pre-measurement uncertainties
$\De A$ and $\De B$, we apply Robertson's relation
to all terms in the left-hand-side of 
\Eq{commutation2}.
Then we obtain the {\em
generalized uncertainty relation for joint measurements}
\begin{equation}\label{eq:GUR}
\ep(A)\ep(B)+\ep(A)\,\De B+\De A\,\ep(B)
\ge \frac{1}{2}|\bracket{\ps|[A,B]|\ps}|.
\end{equation}
From the above, if $\Pi(x,y)$ precisely measures $A$, we have
\begin{equation}\label{eq:noiseless}
\De A\,\ep(B)\ge\frac{1}{2}|\bracket{\ps|[A,B]|\ps}|.
\end{equation}

Now, we shall discuss how the above relations generalize
Heisenberg's relation and related relations obtained previously
\cite{AK65,YH86,91QU,Ish91}.
We say that the joint POVM $\Pi(x,y)$ has 
{\em statistically independent noise} for $A$, 
if the mean noise $\bracket{N_{A}}$
does not depend on the input state $\ps$, or equivalently, if the mean noise
operator $n_{A}$ is a constant operator, i.e., $n_{A}=r I$ for some
$r\in\R$; in this case, we have $n_{A}=\bracket{N_{A}}I$.  
Since the mean noise operator $n(N_{A})$ can be interpreted as
the conditional expectation of $N_{A}$ 
conditional upon the state of the object system, 
the above definition is
consistent with a criterion of the statistical independence requiring that the
conditional expectation be constant. 
In fact, if $\Pi(x,y)$ has statistically
independent noise for $A$,  
any observables $X$ of the object and $N_{A}$
are  statistically independent in the sense that 
\begin{equation}
\bracket{\tX N_{A}}=\bracket{N_{A}\tX}
=\bracket{\ps|X|\ps}\bracket{N_{A}}
\end{equation}
for any state $\ps$ of the object;
the above relations follow from
$\bracket{\tX N_{A}}=\bracket{X\ps|\bracket{\xi|N_{A}|\xi}|\ps}
=\bracket{X\ps|n_{A}|\ps}
=\bracket{\ps|X|\ps}\bracket{N_{A}}$
and $\bracket{N_{A}\tX}=\bracket{\tX N_{A}}^{*}
=\bracket{\ps|X|\ps}\bracket{N_{A}}$.

We say that the joint POVM $\Pi(x,y)$ makes an {\em unbiased
measurement} of  $A$, if the mean output $\sum_{x}\,x\,\Pr\{\bx=x\|\ps\}$
is equal to the
mean $\bracket{\ps|A|\ps}$
of the observable $A$ in any input state.
The above condition is equivalent to the relation
$
\sum_{x}\,x\,\Pi^{A}(x)=A.
$
From \Eq{output-mean}, this is the case
if and only if the mean noise operator vanishes, i.e., $n_{A}=0$,
so that if $\Pi(x,y)$ makes an unbiased measurement of  $A$, then
$\Pi(x,y)$ has statistically independent noise for $A$.
For the $B$ measurement, the corresponding definitions on statistical
independent noise and unbiased measurements are introduced
analogously.

Since the relations $n_{A}=r I$ and $n_{B}=r' I$  obviously imply
$[n_{A},B]=[A,n_{B}]=0$, and hence by \Eq{UVUR2} we conclude 
the following:
{\em If the joint POVM $\Pi(x,y)$ has statistically independent noises for
both
$A$ and $B$, then we have
\begin{equation}\label{eq:Heisenberg-NAB}
\ep(A)\ep(B)\ge
\De N_{A}\De N_{B}\ge\frac{1}{2}|\bracket{\ps|[A,B]|\ps}|
\end{equation}
for any state $\ps$, so that $\Pi(x,y)$ satisfies Heisenberg's relation.}

The above relations were previously proven for the unbiased case 
in Refs.~\cite{91QU,Ish91}.

The standard deviation $\De\bx$ of the output $\bx$ in the state $\ps$
is given by
\begin{equation}
\De\bx
=(\bracket{\ps|O^{(2)}(\Pi^{A})|\ps}
-\bracket{\ps|O(\Pi^{A})|\ps}^{2})^{1/2}.
\end{equation}
Then, for the ancilla $(\cK,\xi,C,D)$ we have
$\De\bx=\De C=\De [\tA+N_{A}]$.  
Thus, if $\Pi(x,y)$ has statistically independent noise for both $A$
and $B$, we have 
\begin{eqnarray}
(\De\bx)^{2}&=&(\De A)^{2}+(\De N_{A})^{2}\ge 2\De A\De N_{A},
\\
(\De\by)^{2}&=&(\De B)^{2}+(\De N_{B})^{2}\ge 2\De B\De N_{B},
\end{eqnarray}
and hence apply \Eq{Robertson} and \Eq{Heisenberg-NAB}
to the product of the above two inequalities, 
we have
\begin{equation}\label{eq:030819d}
\De\bx\De\by\ge|\bracket{\ps|[A,B]|\ps}|.
\end{equation}
The above relation has been previously proven for the unbiased case 
in Ref.~\cite{AG88}.
Relation~\eq{030819d} shows that the uncertainties of the outputs 
increase by the additive fluctuations from the independent
noises, and so that the lower bound for the uncertainty product 
of the outputs is twice as much as that for the measured observables
before the measurement.
However, the relation $[C,D]=0$ suggests that the dependent
noise can reduce the produce $\De\bx\De\by$ arbitrarily small.

\section{Concluding remarks}

In this Letter, we have shown that every joint measurement satisfies
the relations
\begin{eqnarray}
\ep(A)\ep(B)+\ep(A)\,\De B+\De A\,\ep(B)
&\ge&
\ep(A)\ep(B)
+\frac{1}{2}|\bracket{\ps|[n_{A},B]|\ps}|
+\frac{1}{2}|\bracket{\ps|[A,n_{B}]|\ps}|
\label{eq:start}\\
&\ge&
\De N_{A}\,\De N_{B}
+\frac{1}{2}|\bracket{\ps|[n_{A},B]|\ps}|
+\frac{1}{2}|\bracket{\ps|[A,n_{B}]|\ps}|\\
&\ge& \frac{1}{2}|\bracket{\ps|[A,B]|\ps}|.
\end{eqnarray}
If the apparatus has statistically independent noises for
both $A$ and $B$, then we have
$\bracket{\ps|[n_{A},B]|\ps}=\bracket{\ps|[A,n_{B}]|\ps}=0$,
so that the above relations are reduced to Heisenberg's relation
\begin{eqnarray}
\ep(A)\ep(B)
\ge
\De N_{A}\,\De N_{B}
\ge \frac{1}{2}|\bracket{\ps|[A,B]|\ps}|.
\end{eqnarray}
In this case, the product of the output standard deviations
has the twice as much lower bound as 
the input standard deviations, i.e., 
\begin{equation}
\De\bx\De\by\ge|\bracket{\ps|[A,B]|\ps}|.
\end{equation}
In Heisenberg's relation a precise measurement of one observable 
excludes the measurement of the other noncommuting observable
even with any finite noise.
However, from $\ep(A)=0$ our relation concludes 
\begin{equation}\label{eq:end}\\
\De A\,\ep(B)\ge\frac{1}{2}|\bracket{\ps|[A,B]|\ps}|,
\end{equation}
which suggests the possibility that we can overcome the
Heisenberg's limitation of the measurement when one of the 
noises are not statistically independent.

In the preceding discussions, we assume that $\cH$ is finite dimensional
and $\bx$ and $\by$ take only finite number of values.
However, the generalization to infinite dimensional case allowing
continuous valued $\bx,\by$ is rather straightforward.
In the general case, the POVM is defined by any family 
$\{\Pi(x,y)|\ x,y\in\R)\}$ of positive
operators on $\cH$ satisfying  (i) $0\le \Pi(x,y)\le \Pi(x',y')\le I$ if $x\le x'$ and
$y\le y'$, (ii) $\lim_{x\to -\infty,y\to -\infty}\Pi(x,y)=0$,
(iii) $\lim_{x\to \infty,y\to \infty}\Pi(x,y)=I$, and
(iv) $\lim_{x\to a+0,y\to b+0}\Pi(x,y)=\Pi(a,b)$ for any $a,b\in\R$.
Then, the joint probability of obtaining the outcomes 
$\bx\le x$ and $\by\le y$ in the state $\ps$ is described by
$ 
\Pr\{\bx\le x,\by\le y\|\ps\}=\bracket{\ps|\Pi(x,y)|\ps}.
$ 
Then, the marginal POVMs, and moment operators are defined by 
$\Pi^{A}(x)=\lim_{y\to\infty}\Pi(x,y),$  
$O(\Pi^{A})=\int_{\R}x\,d\Pi^{A}(x)$, 
$O(\Pi^{A})=\int_{\R}x^{2}\,d\Pi^{A}(x)$,
and so on; see Ref.~\cite{91QU} for the detail.
Then, we have the same formulas as Eqs.~\eq{start}--\eq{end},
as long as all the relevant terms are finite. 

In Ref.~\cite{03UVR}, it has been shown that the noise for an observable
$A$ and the disturbance for another observable $B$
in a single output measurement using an apparatus $\bA(\bx)$
can be identified with the two noises of $A$ and $B$ 
in the joint measurement of $A$ and $B$ using apparatus $\bA(\bx,\by)$
obtained by the successive measurement consisting of the measurement
using $\bA(\bx)$ immediately followed by a precise $B$ measurement
using an apparatus $\bA(\by)$. 
In this way, the  noise and disturbance relations previously obtained
in Ref.~\cite{03UVR} have been generalized in the model-independent
formulation to the uncertainty relations
for joint measurements in the present Letter.

Now we return to the joint measurement of $P+P'$ and $Q$
for two independent pairs $(Q,P)$ and $(Q',P')$ of conjugate observables. 
Let $\ps=\ps_{1}\otimes\ps_{2}$.
Since $[P+P',Q]=i\hbar$, from \Eq{GUR}, we generally have
\begin{equation}
\ep(P\!+\!P')\ep(Q)\!+\!\ep(P\!+\!P')\De Q\!+\!\De (P\!+\!P')\ep(Q)
\ge \frac{\hbar}{2}.
\end{equation}
The above relation does not exclude 
the possibility of having
$\ep(P\!+\!P')=0$, and in this case we have
\begin{equation}
\De (P\!+\!P')\ep(Q)\ge \frac{\hbar}{2}.
\end{equation}
Let $\ps=\ps_{1}\otimes\ps_{2}$.  Then, we have
$[\De  (P\!+\!P')]^{2}=(\De P)^{2}+(\De P')^{2}$.
Thus we have
\begin{equation}
\ep(Q)^{2}\ge \frac{\hbar^{2}}{4(\De P)^{2}+4(\De P')^{2}}.
\end{equation}
The above relation gives general limit for the measurement of
position $Q$ without disturbing the total momentum $P+P'$.
For the example in the beginning of the present Letter,
we have $\ep(Q)=\De Q'$, and this would be optimal if $P$ could be
prepared definitely, i.e., $\De P=0$, and $Q$ is prepared in the minimum 
uncertainty state, i.e., $\De Q'\De P' =\hbar/2$.
Therefore, the present results on generalizing Heisenberg's relation 
settles the confusion for the accuracy of measurement in the presence of
a conserved quantity; see \cite{91CP,02CLU}
for relevant discussions.

\section*{Acknowledgements}
This work was supported by the 
Strategic Information and Communications R\&D Promotion Scheme
of the MPHPT of Japan,  by the CREST
project of the JST, and by the Grant-in-Aid for Scientific Research of
the JSPS.

\end{document}